\documentclass[10pt]{amsart} 
\usepackage{graphicx}

\usepackage{caption}
\usepackage{subcaption, wrapfig}

\usepackage{amsmath,amsfonts,amsthm,amssymb,graphics,color, hyperref}
\usepackage{caption, subcaption}
\usepackage{graphicx}
\usepackage[countmax]{subfloat}

\newtheorem{theorem}{Theorem}
\newtheorem{principle}{Principle} 
\newtheorem{defn}{Definition}

\theoremstyle{remark}

\newcommand{\fp}{\mathfrak{p}} 
\numberwithin{equation}{section}

\newcommand{\fS}{\mathfrak{S}} 
\newcommand{\R}{\mathbb{R}}

\renewcommand{\epsilon}{\varepsilon}

        \definecolor{pink}{rgb}{1,0,1}

\begin{document}
\title[Many ways to stay in the game]{Many ways to stay in the game\\  Individual variability maintains high biodiversity in planktonic micro-organisms}

\author[S. Menden-Deuer]{Susanne Menden-Deuer \\ Graduate School of Oceanography \\ University of Rhode Island\\}  

\author[J. Rowlett]{Julie Rowlett \\ Max Planck Institut f\"ur Mathematik, Bonn}\address{J. Rowlett's current address:   Chalmers University and the University of Gothenburg \\ Mathematical Sciences \\ 41296 Gothenburg Sweden}   


\maketitle

\begin{abstract}   
In apparent contradiction to competition theory, the number of known, co-existing plankton species far exceeds their explicable biodiversity - a discrepancy termed the \emph{Paradox of the Plankton.}   We introduce a new game-theoretic model for competing micro-organisms in which one player consists of all organisms of one species.   The stable points for the population dynamics in our model, known as strategic behavior distributions (SBDs), are probability distributions of behaviors across all organisms which imply a stable population of the species as a whole. We find that intra-specific variability is the key characteristic that ultimately allows co-existence because the outcomes of competitions between individuals with variable competitive abilities is unpredictable.  Our simulations based on the theoretical model show that up to 100 species can coexist for at least 10000 generations, and that even small population sizes or species with inferior competitive ability can survive when there is intra-specific variability. In nature, this variability can be observed as niche differentiation, variability in environmental and ecological factors, and variability of individual behaviors or physiology.  Therefore previous specific explanations of the paradox are consistent with and provide specific examples of our suggestion that individual variability is the mechanism which solves the paradox.  
\end{abstract}
 

\section{Introduction} 
Planktonic organisms are integral to the functioning of the global ecosystem, generating 50 \% of the organic C and breathable O$_2$ \cite{fields}, fuelling fisheries, and serving as key agents in major biogeochemical cycles \cite{falk08}.  Understanding and predicting the abundance, distribution, and diversity of plankton is critical to predicting their globally important geochemical footprint and the effects of changing environmental conditions. 

The number of coexisting planktonic species \cite{falk04} far exceeds the expected and explicable number based on competition theory \cite{har60}. This has been termed \emph{ the Paradox  of the Plankton}  \cite{hut61}. There have been important contributions demonstrating that specific factors can enhance coexistence of multiple species (e.g. tradeoffs \cite{til}, competition for multiple resources \cite{hw99}, chaotic fluid motion \cite{chaos}, localized rather than regional competition \cite{kerr}).  However, these conditions are not met all of the time and are not inherent to the species, thus not subject to natural selection. 

There is evidence for high variability in a range of physiological, demographic, and morphological traits among phylogenetically distinct microbes and within a single species \cite{fred}. Phenotypic plasticity in important traits is commonly observed in microbial plankton, including in global, inter- and intra specific patterns of temperature regulation  \cite{Thomasetal2012}, \cite{boyd}; and responses to elevated pCO2 concentrations  \cite{Schaumetal2013}. Species even show and retain possibly maladaptive traits, reflecting ancient, rather than current environmental conditions \cite{hutch}.  Plasticity in traits was identified as a key characteristic to adapting to changing or novel conditions and has been associated with elevated success in range expansion of invasive species \cite{Leeetal2003}.  Phototrophic plankton display considerable variability in important traits, including tolerance of environmental conditions \cite{brand}, elemental composition \cite{moal}, and growth rate \cite{kim-md}.  A comprehensive study of intra-strain variability in one phytoplankton species showed distinguishing characteristics in cell size, maximum growth and photosynthesis rates, tolerance of low salinities, resource use, and toxicity \cite{fred}. Molecular analyses discovered that such variability is genetically rooted \cite{ryne}, \cite{whitt}.  Intra-specific variability is further documented in \cite{kremp}.   The maintenance of individual variability may be supported by the complexity of cellular morphology \cite{cell}, \cite{hu01}.  

Motility patterns are also observed to be variable on the individual level; for example a small fraction of a population will move away from a limiting resource or fail to avoid predators \cite{mdg}, \cite{hmd}, \cite{smd2010}. These same observations of individually variable motility patterns appear to be strategic on the population level for example by fleeing predators \cite{hmd} or by using vertical swimming patterns to optimize the acquisition of light and nutrients \cite{kl}.   

We created a new game theoretic model for competing micro-organisms to provide the theoretical foundation for the observed individual variability.  Previous game theoretic models such as \cite{al} assume that interspecies rankings are fixed and identical across all individuals.  This assumption excludes individual variability.  In the lattice version of a multi-species Lotka-Volterra competition model in \cite{mi}, the mathematical equilibrium is the exclusion of all but one superior species.  In that model co-existence is only possible if interactions are local rather than global.  However, interactions in an ocean or lake environment do not remain local, and the experimental data of \cite{kerr} showed that competing species without individual variability which are periodically mixed will cease to co-exist.  

Our new theoretical model for competition among micro-organisms and the resulting population dynamics  incorporates the following key features:  (1)  individual organisms can draw from a distribution of behaviors rather than expressing a static, predetermined behavior;   (2) microbial populations consist of multiple, clonal individuals; (3) the survival of a species is a cumulative function of the success across its individuals.  Our approach based on individual variability is inherent to a species and therefore subject to natural selection.  We find that incorporation of intra-specific variability results in the co-existence of numerous species, even when population sizes are low or competitively inferior behaviors are incorporated.  Thus, we suggest that intra-specific variability is the mechanism that explains the paradox of the plankton.  

This work is organized as follows.  In \S 2, we present the mathematical background and the theoretical model.  This model suggests that if there is variability across individuals, then there is an infinite set of stable points for the population dynamics.  These stable points, known as \em strategic behavior distributions (SBDs) \em represent all the different probability distributions of strategies across individuals which guarantee survival of the species as a whole.  Using a game to model interspecies competition, this abundance of SBDs means that there are  \em many different ways to stay in the game, \em  biologically corresponding to many different ways for species to co-exist, as long as there is some variability among individuals and resources allow.  In case of insufficient resources, this means that there are many different ways for all species to fare equally well.  We tested the theoretical model and its predictions in \S 3 by implementing computer simulations of competing species.  The results show that SBDs support species co-existence in a large and realistic parameter space varying the number of competing species, number of individuals, and simulation durations.  A discussion of how our theory fits with the competitive exclusion principle and relates to previous explanations of the paradox of the plankton comprises \S 4.  We summarize the methods used in our simulations and mathematical proof in \S 5.  

\section{Game-theoretic Model}
\subsection{Game Theory Preliminaries} 
In a classic game theory sense, players are viewed as individuals.  A specific move in a game is known as a \em pure strategy.  \em  If a player has only one pure strategy, then he cannot affect the outcome of competition, so his role is marginalized.  If a player has at least two pure strategies, then a \em mixed strategy \em is a probability distribution across that player's pure strategies.  Take for example the Rock-Paper-Scissors (RPS) game.  A mixed strategy is a list of three numbers between zero and one which add up to one.  These numbers are the probability of executing Rock, Paper, or Scissors.  

We extended the definition of a player as consisting of many individuals all belonging to a single species.  An $n$-player game models competition between $n$ species. In our model, a mixed strategy for a player (which we define to be a whole species), is a probability distribution of that player's pure strategies across its individuals.  For example, in RPS, this would be the probability that a randomly selected individual draws Rock, or Paper, or Scissors.  Each player has an associated payoff function ($\wp_i$) for $i=1, \ldots, n$ which depends on the (mixed) strategies of all players.  

To model competition amongst different species of micro-organisms we use an $n$-player non-cooperative game.  Such a game is based on the absence of coalitions or communication between players.  The $i^{th}$ player has a set of $m_i$ pure strategies, each of which can be identified with one of the standard unit vectors in $\R^{m_i}$.  The set of mixed strategies corresponds to a probability distribution over pure strategies and can be identified with the convex subset 
$$\fS_i \cong \left\{ x = (x_1, \ldots, x_{m_i}) \in \R^{m_i} : x_j \geq 0 \forall j, \quad \sum_{j=1} ^{m_i} x_j =1 \right\}.$$
The total strategy space over all players can then be identified with 
$$\fS \cong \prod_{i=1} ^n \fS_i \subset \R^{N}, \quad N = \sum_{i=1} ^n m_i.$$
Each player has an associated payoff function $\wp_i : \fS \to \R$.  The payoff functions are linear in the strategy of the respective player.  That is, if all other players' strategies are fixed, the payoff function is a linear function from $\fS_i \to \R$.  Typically, the payoff function is assumed to be continuous on $\fS$.  In \cite{nash}, Nash proved that for every such game, there exists at least one ``equilibrium strategy,'' which in the following sense is the ``best'' strategy for all players.  For $s \in \fS$, let $\wp_i (s; \sigma; i)$ denote the payoff for the strategy in which the $i^{th}$ player's strategy according to $s$ is replaced by $\sigma \in \fS_i$.  An equilibrium strategy $s$ satisfies 
$$\wp_i (s) \geq \wp_i (s; \sigma_i; i) \quad \forall \sigma_i \in \fS_i, \quad \forall i=1, \ldots, n.$$
In other words, if a single player changes his strategy, he cannot increase his payoff.  

\begin{theorem}[Nash \cite{nash}] For any $n$-person non-cooperative game such that the payoff functions are linear in the strategy of each player and are continuous functions on the total strategy space there exists at least one equilibrium strategy. 
\end{theorem} 

Although an equilibrium strategy is in a certain sense the best possible strategy, there may be many other strategies such that the corresponding payoff to all players is \em identical \em to their payoffs according to an equilibrium strategy.  This means that the \em feedback \em for such strategies is \em identical \em to the feedback for the equilibrium strategy. We propose to consider the level sets of the total payoff function, that is the sets of strategies so that each player has constant feedback within this set of strategies.  This means that varying strategies within a level set of the total payoff function doesn't change the payoff to any of the players, so if the players' strategies vary within this set, the resulting feedback is unaffected.  It turns out that these level sets of the total payoff function are in general quite large.    

\subsection{Population Dynamics}
The success or failure of micro-organisms to acquire necessary resources can be identified by the rate of population increase or decrease.  So, we propose to use the payoffs in a game theoretic model to define the population dynamics:  positive payoff yields population increase, negative payoff yields population decrease.  A \em strategy \em for a species is a probability distribution of behaviors across all individuals of that species.  The payoff is a cumulative function measuring the average success over all individuals, so the change in population should also be proportional to the current population.  This is how we came to define the differential equation for the population dynamics of the $i^{th}$ species,  
\begin{equation} \label{rep-eq} \fp_i '(t) = \fp_i (t)  \wp_i (s). \end{equation}  
Above $\fp_i(t)$ is the population of the $i^{th}$ species at time $t$, $\fp_i'(t)$ is the instantaneous change in the population at time $t$, and $\wp_i(s)$ is the payoff to the $i^{th}$ player for the game according to the strategy $s$ over all players.  

\begin{defn} 
A strategic behavior distribution (SBD) is a rest point for the equation (\ref{rep-eq}), that is a strategy $s$ such that $\wp_i (s) = 0$, for each $i=1$, $2$, $\ldots$, $n$.  The set of all SBDs is therefore the zero-level set of the total payoff function.  
\end{defn} 

The novelty here is our application of geometric measure theory to game theory and subsequently to the population dynamics of micro-organisms.  The mathematical core of this work is the following result.  

\begin{theorem} For an $n$-player game, assume that each player has at least two pure strategies and that the payoff function is Lipschitz continuous.  Let $N$ denote the total number of pure strategies across all players.  Assume at least one of the following holds:  (1) the game is zero-sum and/or (2) at least one player has three or more pure strategies.  If (1) holds, then let $k = N-2n +1$; if only (2) holds then let $k=N-2n$.  With respect to $k$-dimensional Hausdorff measure, the level sets of the total payoff function $\wp := (\wp_1, \wp_2, \ldots, \wp_n)$ $k$-almost always have positive $k$-dimensional Hausdorff measure.  
\end{theorem} 

The proof of the theorem is provided in \S 5.  The following principle is the biological interpretation of our mathematical theorem.  

\begin{principle} 
If each species of competing micro-organisms has variability across individuals, then if there exists one SBD, there is in a generic sense (almost always) an infinite set of SBDs.  Moreover, for a specific set of respective feedbacks to all species, there are almost always infinitely many different behavior distributions over the individuals of each species which yield those same respective feedbacks to each species.  
\end{principle} 

This principle is depicted in Figures \ref{rps} and \ref{fig1}, which show that although an equilibrium strategy may be a single point, there is an entire \em triangular region \em or \em line, \em respectively, containing infinitely many different strategies which all give feedback to the competing species which is \em identical \em to their feedback at the equilibrium point.  We propose that the large level sets of the total payoff function support the large biodiversity observed in planktonic microbes.

\section{Model Simulations}
We implemented computer simulations of competing species using a symmetric, two-player, zero-sum game with two pure strategies, W and L, such that W (win) dominates L (lose) (see Fig.~ \ref{fig1}).   In a natural environment, competition occurs between individual organisms rather than whole populations.  For the theorem and our general principle, we note that games are neither required to be symmetric nor zero-sum; we made this assumption for computation convenience.  

At each round of competition, each individual was randomly assigned to compete with an individual belonging to a different species. The individual's competitive ability was randomly assigned, represented by a number between 0 and 1, selected from the particular SBD for its species (Fig. ~\ref{fig2}).  The three possible outcomes were (1) draw: identical values for both individuals, (2) win: larger value than competitor, or (3) lose: lower value than competitor.  A draw has both individuals remaining, a win results in the doubling of that individual, whereas a lose results in the removal of that individual.  To assess species co-existence these competitions were repeated for a number of individuals, species and durations, as specified below.  Species survival was then assessed across all its individuals.  

\subsection{Model simulation results}
\begin{figure*}
        \centering
        \begin{subfigure}[b]{0.5\textwidth}
                \includegraphics[width=\textwidth]{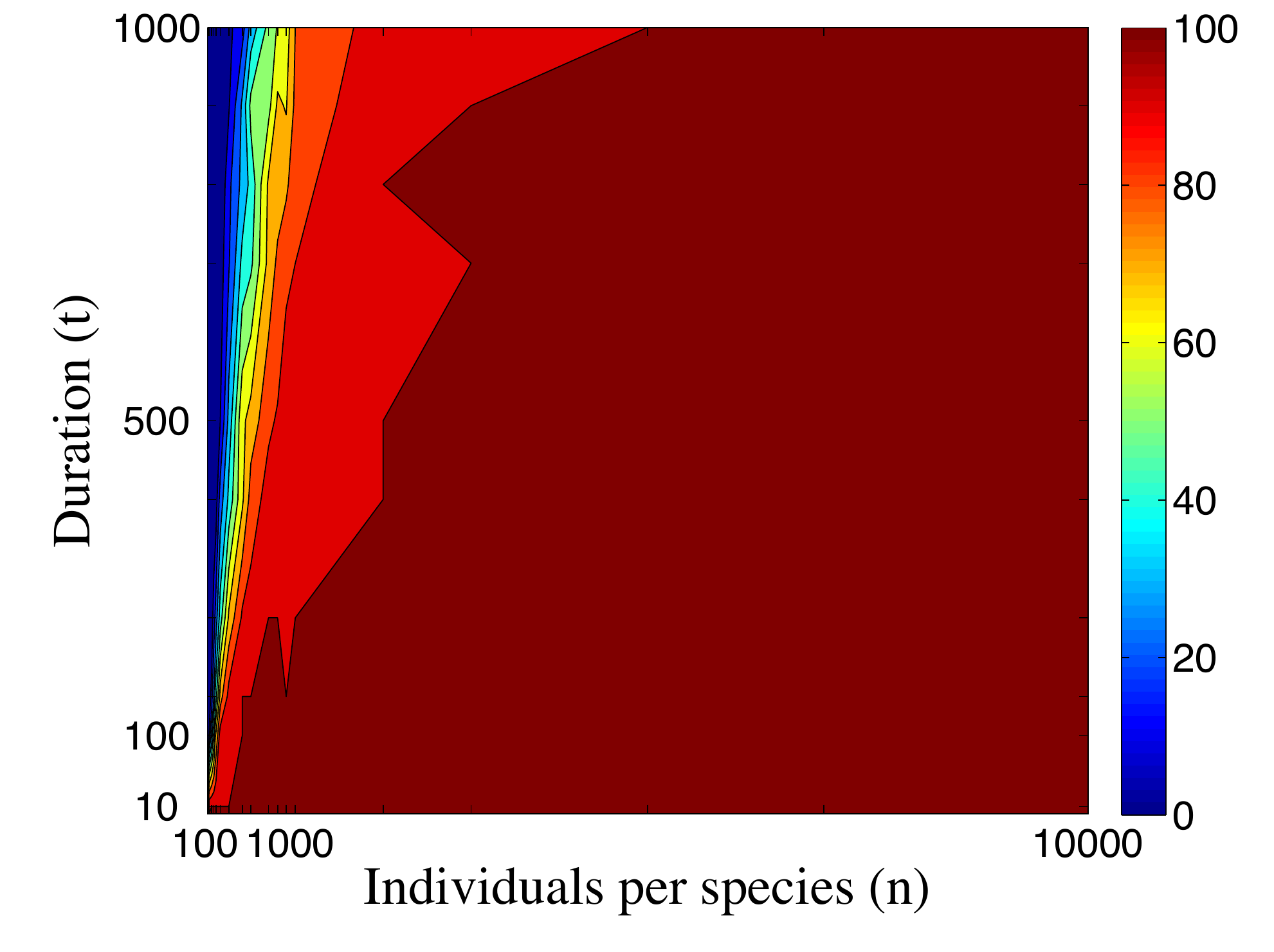}   
                \caption{}
        \end{subfigure}%
        \begin{subfigure}[b]{0.45\textwidth}
                \includegraphics[width=\textwidth]{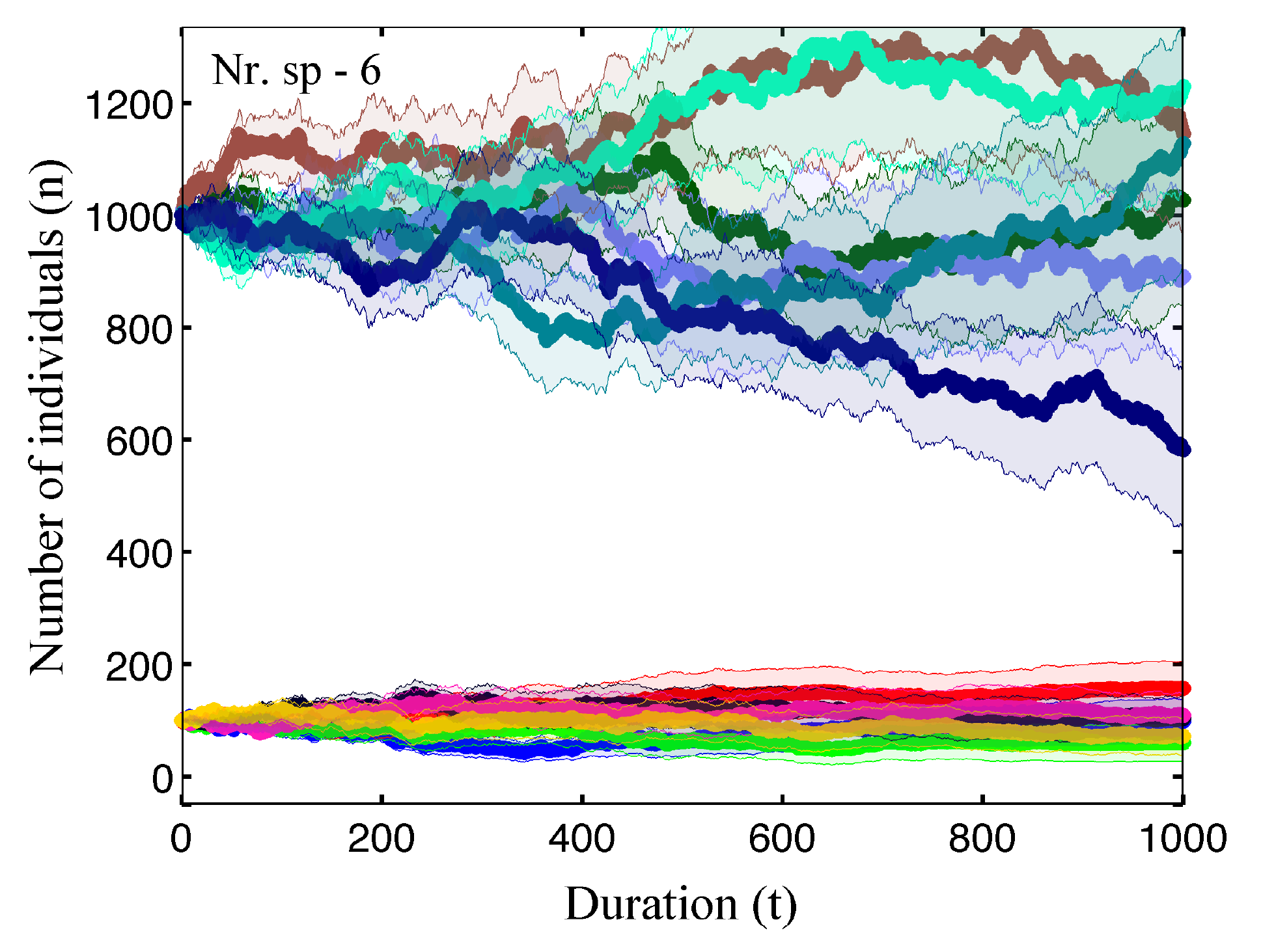} 
                \caption{}
                \label{6sp}
        \end{subfigure}
        
                \begin{subfigure}[b]{0.45\textwidth}
                \includegraphics[width=\textwidth]{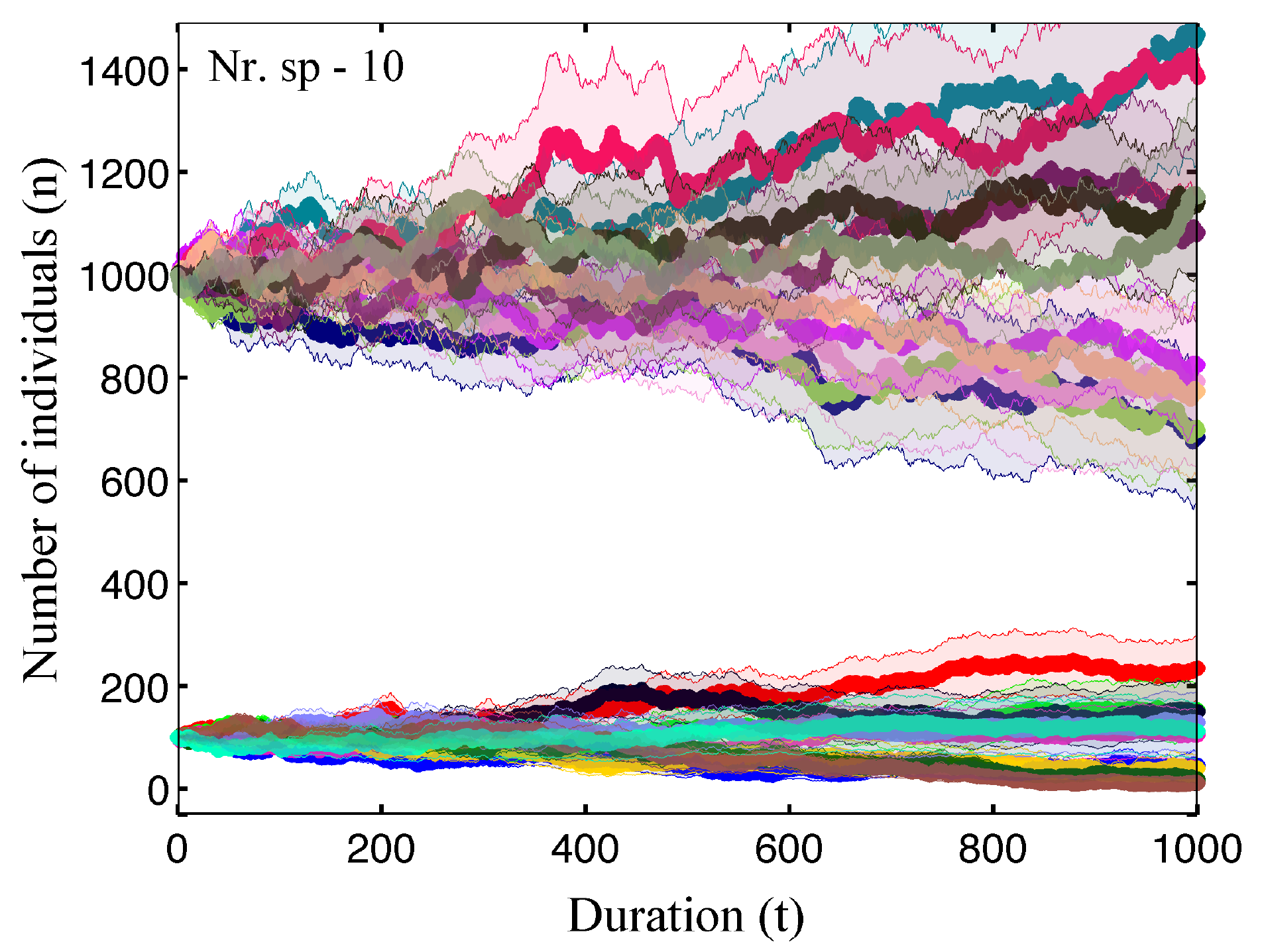}
                \caption{}
                \label{10sp}
        \end{subfigure}    
               \begin{subfigure}[b]{0.45\textwidth}
                \includegraphics[width=\textwidth]{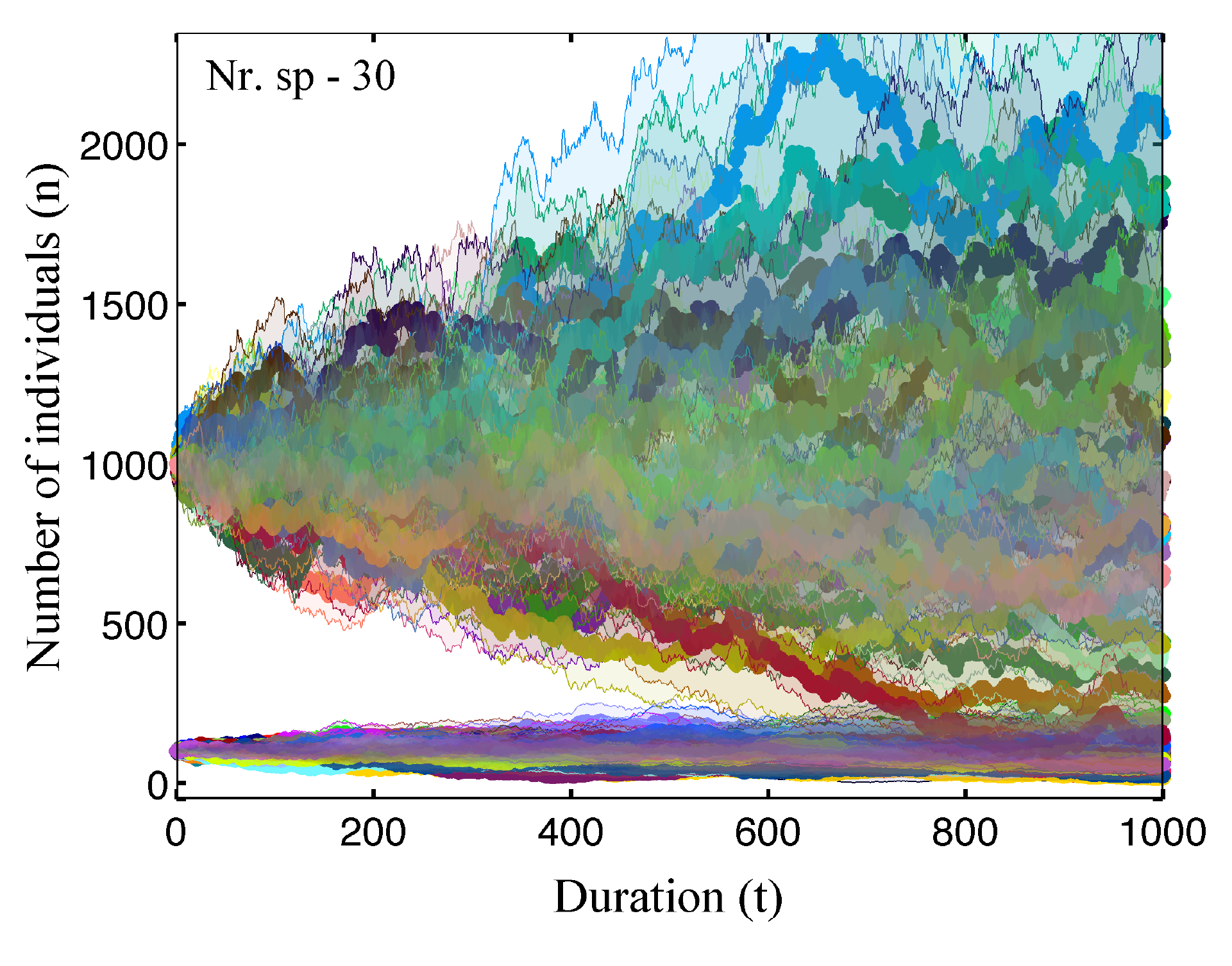}
                \caption{}
                \label{30sp}
        \end{subfigure}
        
               \begin{subfigure}[b]{0.5\textwidth}
                \includegraphics[width=\textwidth]{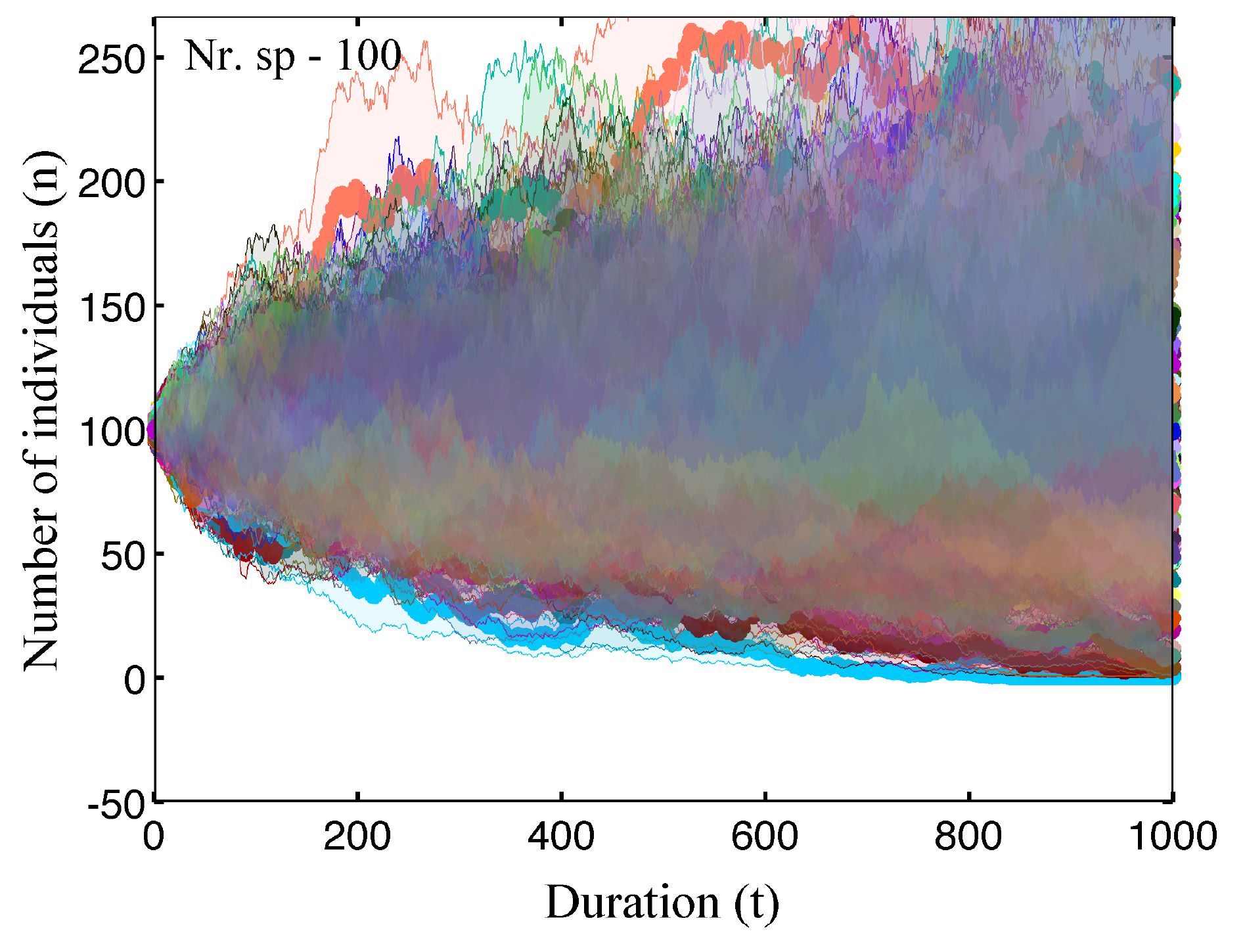}
                \caption{}
        \end{subfigure}
    \caption{The outcomes of the competition model.  (A) shows the mean over 100 repetitions of the randomized model simulation of two competing species with the vertical axis showing the averaged percent probability that two species co-exist.  Probability of coexistence of both species increased with simulation duration and individuals per species.  Co-existence for at least 1000 generations was observed in at least 90\% of the simulations when there were initially at least 1000 individuals per species. (B), (C), (D), and (E) show the number of individuals for 6, 10, 30, or 100 competing species, respectively, averaged over 30 repetitions.  Initial starting abundances were 10, 100, or 1000 individuals per species.  The shaded areas show the variance amongst runs. No extinctions were observed.}
\label{fig3} 
\end{figure*}

In tournaments matching 2 species, persistence of both species was observed when each species was characterized by an SBD (Fig.~ \ref{fig2}). Although extinctions did occur due to stochastic fluctuations when population size was small ($\lesssim 100$ individuals), persistence of 2 species was consistently observed over dozens of generations at larger population sizes (Fig.~ \ref{fig3} (A)). Our model accurately reproduced rapid (within $< 10$ generations) species extinction when individual competitive abilities were identical across the entire species, consistent with the competitive exclusion principle.  

Extension of the simulations to include up to 100 species showed that consistent persistence for 1000 generations was observed with population sizes of 500 or more individuals (Fig.~ \ref{fig3} (B)--(E)). Irrespective of the initial number of species,  approximately $100$ individuals were sufficient for total species diversity to persist in at least 80\% of the replicated runs. Competitive exclusion, that is, total extinction of all but one species was only observed when there were fewer than 10 individuals per species. Species persistence did not imply lack of variation in relative abundance. Depending on the number of species and individuals, there were considerable fluctuations in the abundance, and some species could be rare, but they persisted for at least 1000 competitions. 

The upper limits of the simulation durations, number of species, and population sizes were determined by the availability of CPUs due to the computationally intensive nature of modelling individual-based interactions. However, while technically challenging, simulating larger population sizes would only further support the proposed ideas. Larger population sizes would provide an even greater buffer against extinction; the probability of extinction decreases with increasing population size, as shown in Figure 1A. It is therefore not implied that extinctions would be observed when more species and/or individuals are evaluated simultaneously but in fact that extinctions would be even more unlikely.  Simulation of population sizes of $10^4$ individuals provides an example of probable cell-cell encounters within a time frame (minutes to hours) relevant to individual cells, within a volume of sub-ml to mls.

\subsection{Lattice Grid Simulations}                                                                                                                                                                                                                                                                                                                                                                                                                                                                                                                                                                                                                                                                                                                                                                                                                                                                                                                                                                                                                                                                                                                                                                                                                                                                                                                                                                                                                                                                                                                                                                                                                                                                                                                                                                                                                                                                                                                       
Landscape ecology has identified the importance of spatial variability in maintaining species diversity \cite{l92}.  Restricting competition to proximate competitors has been shown to maintain species diversity in microbes \cite{kerr}. To examine the effect of spatial structure in maintaining species diversity, we formulated a spatially explicit lattice model following RPS dynamics. Spatial variability was mimicked by allowing species to compete either locally or globally, that is either only with  individuals directly adjacent or throughout the entire matrix.  Individual variability was mimicked by allowing individuals of each of three species to vary the strategy in each step by selecting either Rock, Paper, or Scissors.  Such variability in behavior has been shown, for example in sensitivity, resistance, and toxicity of bacterial strains \cite{cz}.  Constant behavior types corresponded to three species:  one species whose individuals always drew Rock, a species whose individuals always drew Scissors, and a species whose individuals always drew Paper.  

               \begin{figure}
        \centering
        \begin{subfigure}[b]{0.4\textwidth}
                \includegraphics[width=\textwidth]{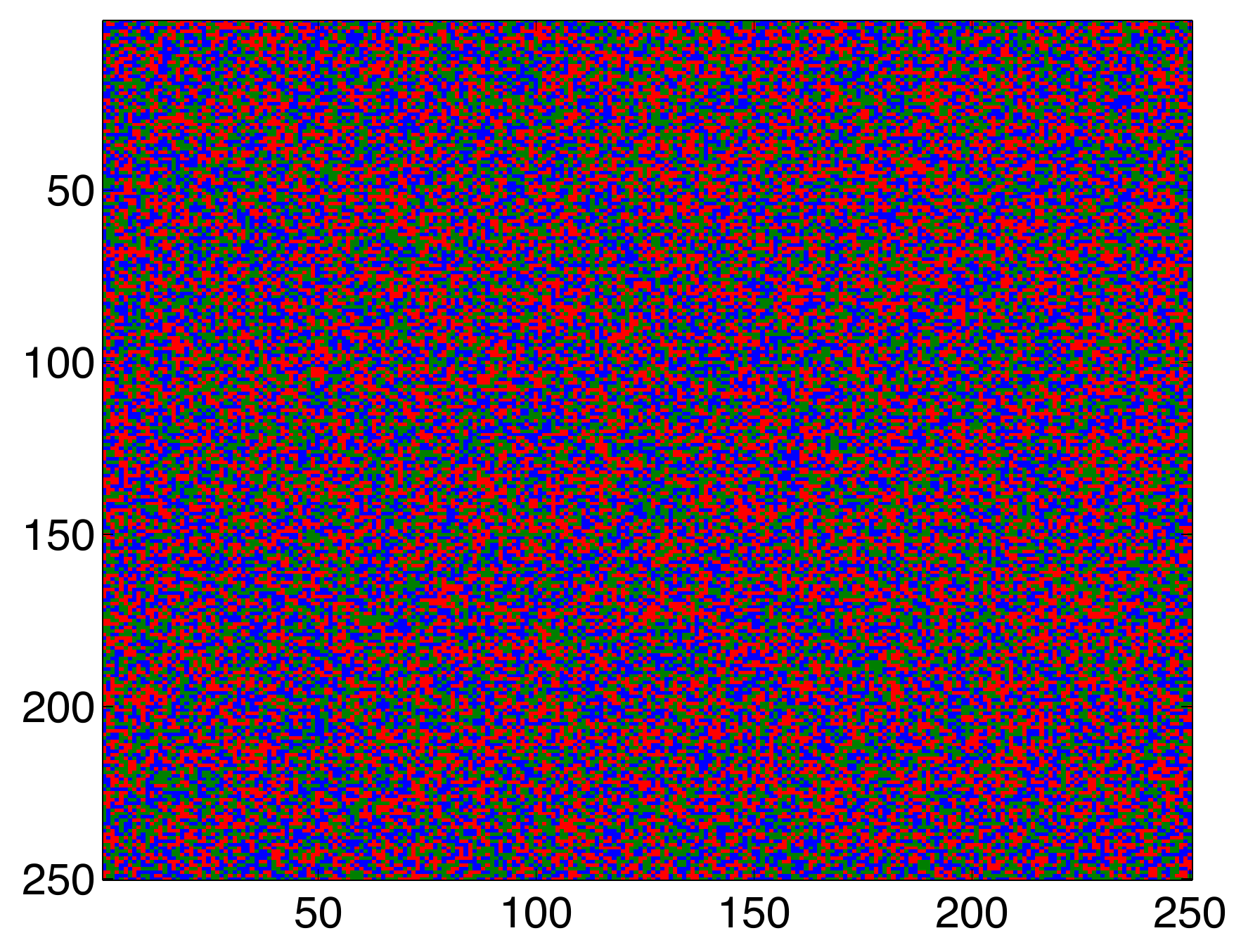}  
                \caption{}
                \label{whole}
        \end{subfigure}%
        \begin{subfigure}[b]{0.4\textwidth}
                \includegraphics[width=\textwidth]{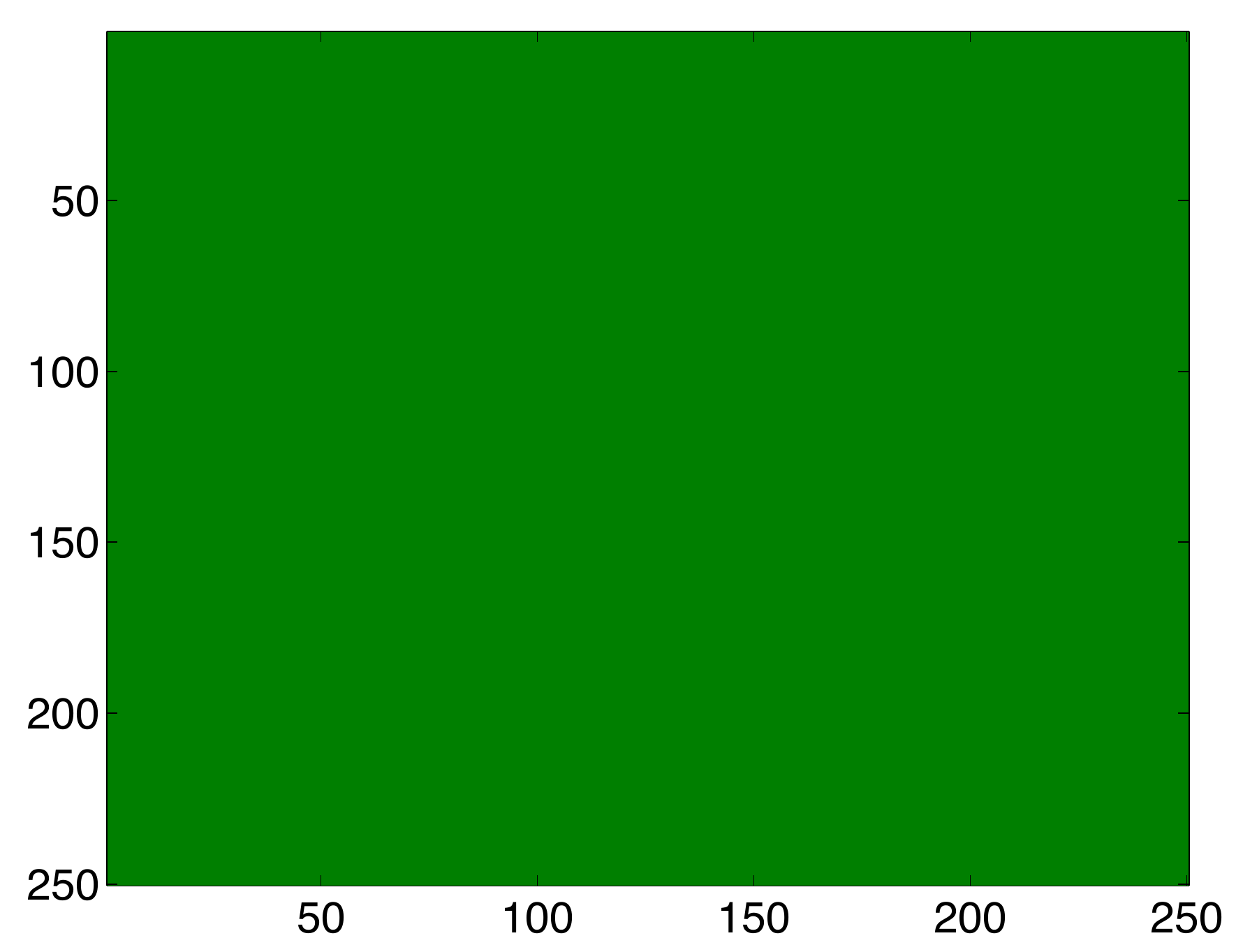}
                \caption{}
                \label{center}
        \end{subfigure}    
        
                  \begin{subfigure}[b]{0.4\textwidth}
                \includegraphics[width=\textwidth]{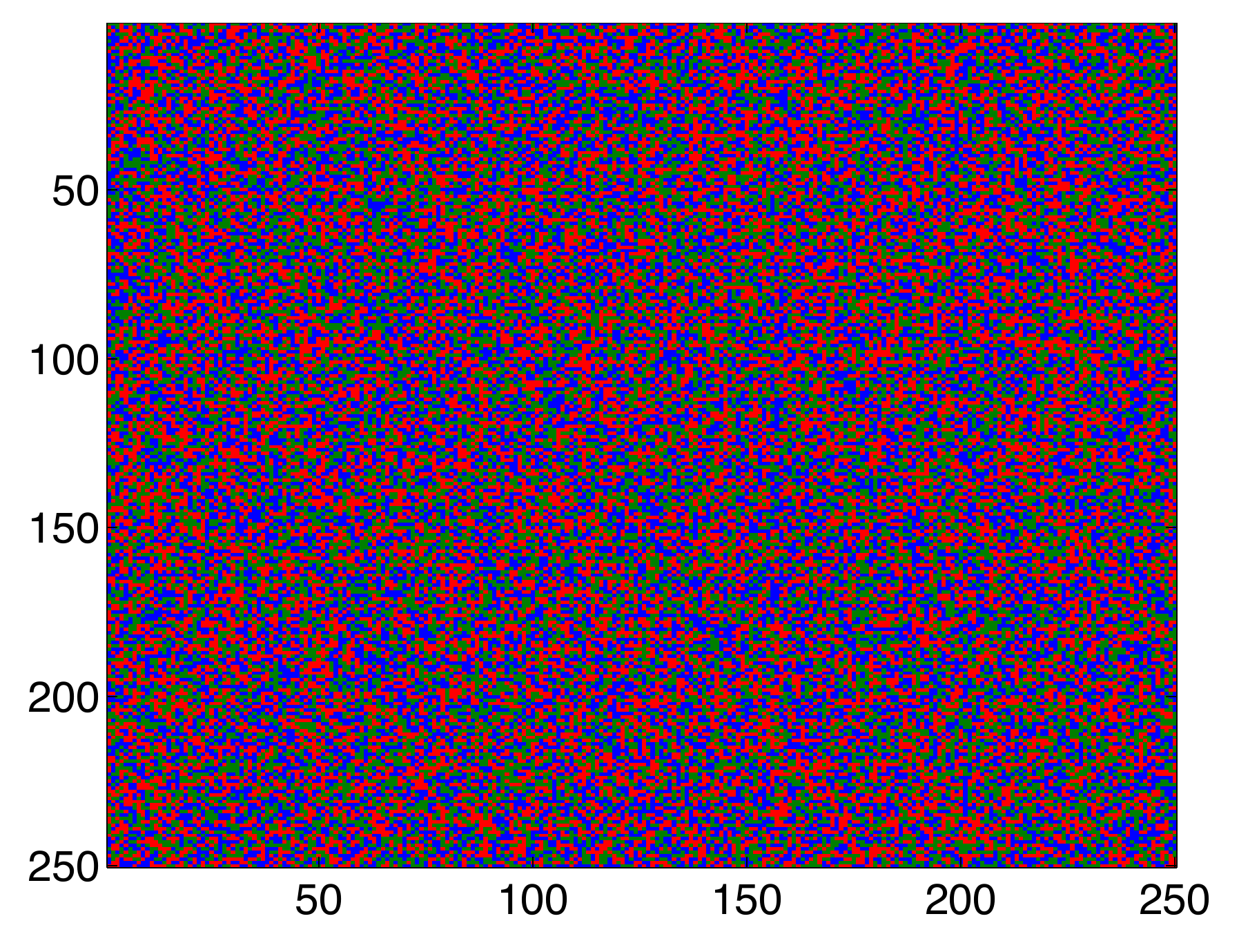} 
                \caption{}
                \label{edges}
        \end{subfigure}
                \begin{subfigure}[b]{0.4\textwidth}
                \includegraphics[width=\textwidth]{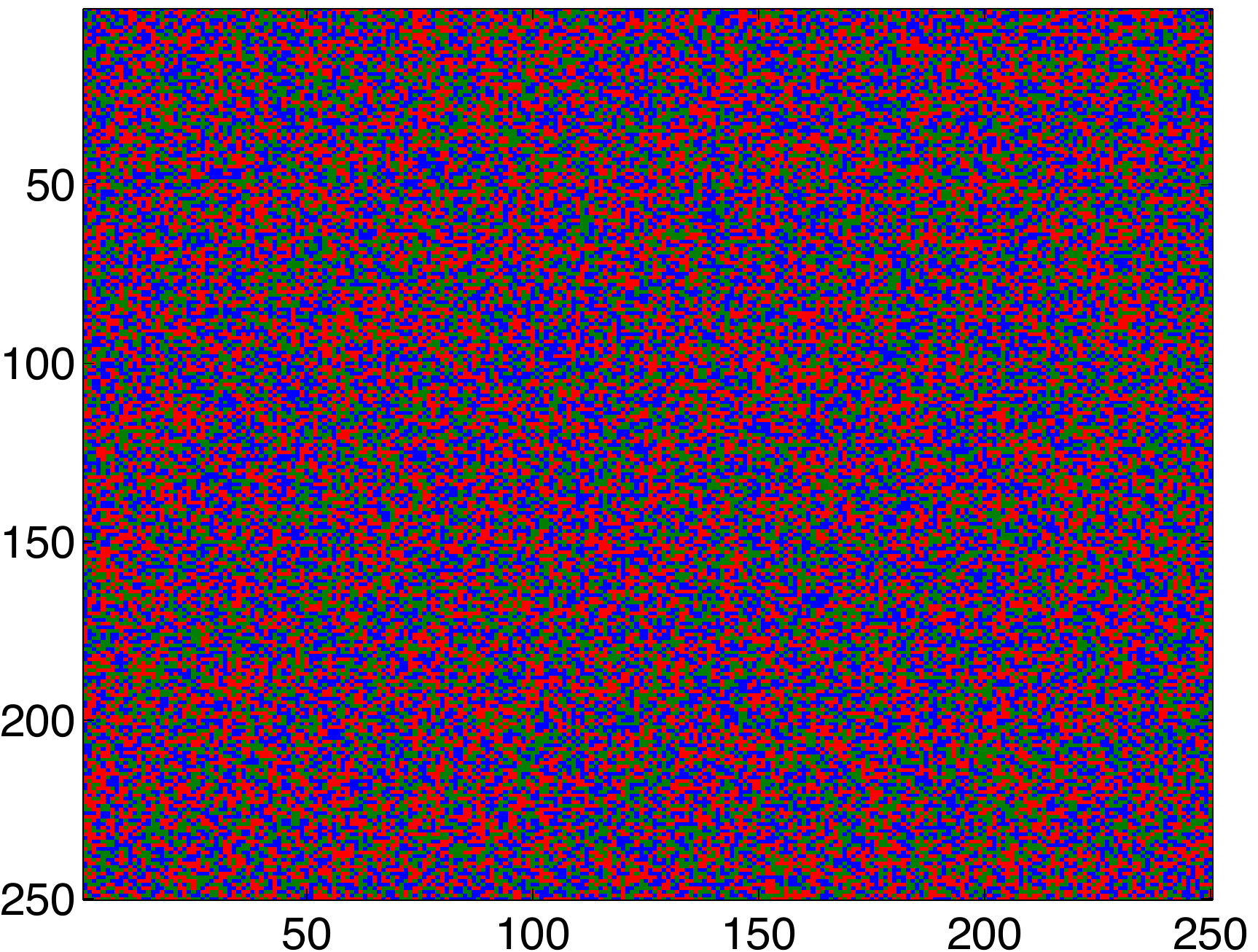}
                \caption{}
                \label{inferior}
        \end{subfigure}
        
        \begin{subfigure}[b]{0.6\textwidth}
                \includegraphics[width=\textwidth]{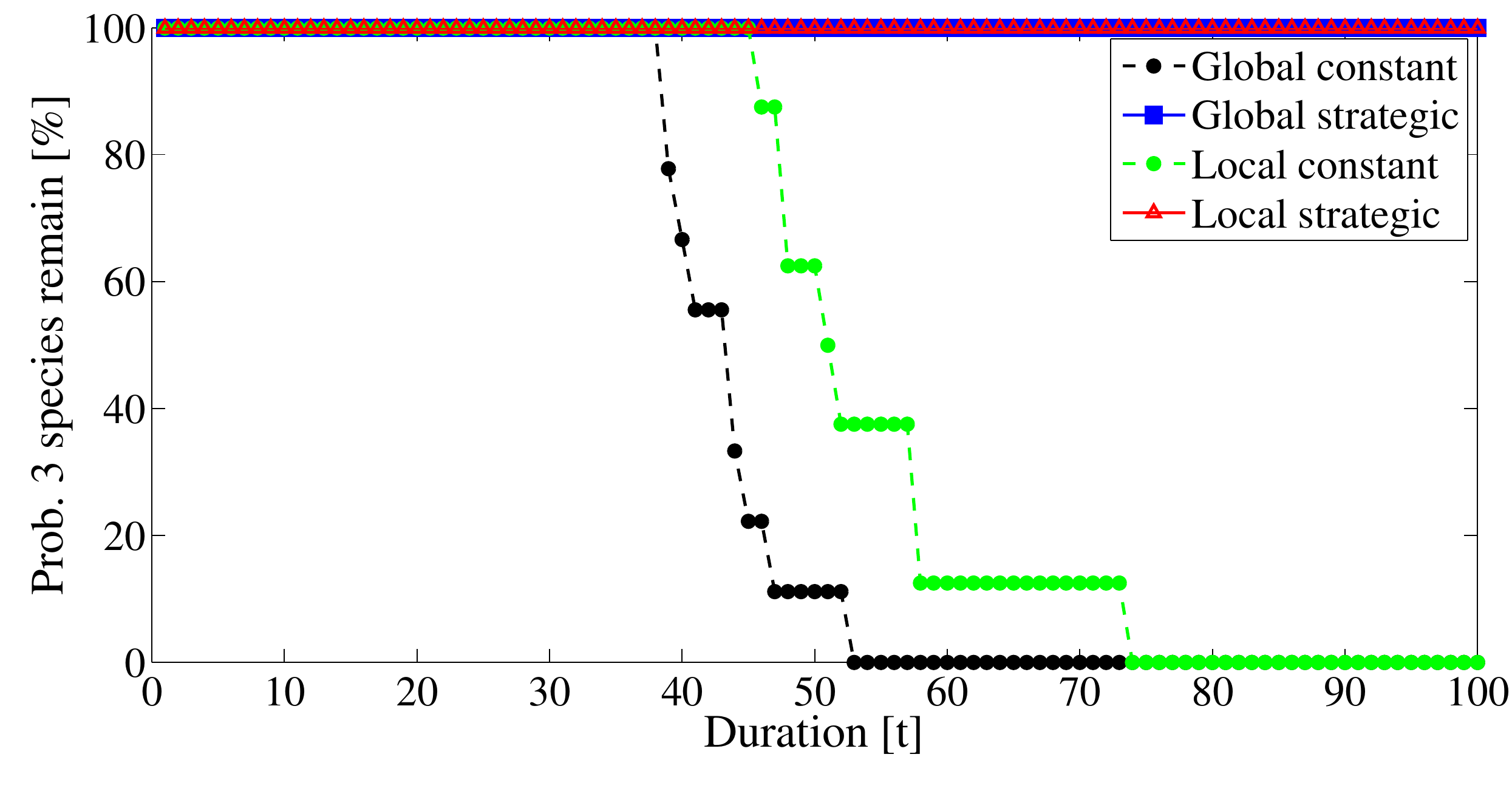}
                \caption{}
                \label{edges}
        \end{subfigure}    
    \caption{The spatially explicit RPS competition model.  (A) shows the initial state of the lattice model with three species colored red, blue, and green.  (B) and (C) show the lattice after 100 rounds of global competition with constant and variable strategies, respectively.  (D) shows the lattice after 100 rounds of local competition with variable strategies.  (E) shows the probabilities over time, averaged over 30 simulations, that three species remain in coexistence.}
\label{fig4} 
        \end{figure}

Our model confirmed the results of \cite{kerr}, that when each species expresses constant behaviors, local competition maintains species diversity longer than global competition (Fig.~ \ref{fig4}). Irrespective of whether competition was global or local, no species extinctions were observed when individuals had variable strategies  (Fig.~ \ref{fig4}).

\section{Discussion}
Representing species through many individuals with variable competitive abilities brought inter-species competition from an aggregate, species level to the individual level.  Our results show that high species diversity is explicable when the underlying behavior distributions allow for variability across individuals, and conversely, that SBDs maintain species diversity.  This variability across individuals may be intrinsic, corresponding to inherent behavioral heterogeneity, as well as extrinsic, resulting from environmental, ecological, spatial, or temporal variability.  It is of course well known that \em identical \em species can co-exist in a neutral sense.  The key here is that the large (infinite) number of \em different \em SBDs allow many \em different \em species to co-exist in a neutral sense because these SBDs imply \em many different ways to stay in the game.  \em  Our model simulations showing prolonged co-existence based on a simple game imply the same results for more complicated games and payoff functions because the more complicated the game, and the more variability across individuals, the larger the set of SBDs.  This can be seen for example in Figures \ref{rps} and \ref{fig1}.  In the RPS game players each have three pure strategies, whereas in the game in \ref{fig1}, players each have only two pure strategies.  Consequently, RPS is a more complicated game, and correspondingly the set of SBDs for the RPS is a whole triangular region (a two-dimensional set), which is much larger than the set of SBDs in \ref{fig1}, which is a line segment (a one-dimensional set).

\subsection{Previous explanations for the paradox} 
There have been multiple plausible explanations for the paradox of the plankton, typically focused on a single factor such as resource competition.  However, organisms simultaneously compete not just for resources, but also require defence from predation and are subject to multiple stressors and selection factors.  

One explanation proposed to resolve the paradox is niche differentiation.  If species compete for several resources and each possesses a ranking in terms of the success of that species at obtaining each resource, competition can be modelled using a competitive network, like a tournament.  This was the approach taken in \cite{al}.  However, with  this approach, they showed that an even number of species can never co-exist.  They assumed that all individuals of a single species have identical, fixed, competitive abilities.  For example, if species 1 is better at obtaining resource A while species 2 is better at obtaining resource B, then \em every \em individual in species 1 is better at obtaining resource A, and \em every \em individual in species 2 is better at obtaining resource B.  This assumption is not actually required for niche differentiation.  Different species can be better or worse on average at obtaining specific resources yet variable on the individual level.  Moreover the success of individuals may (and likely does) fluctuate over time.  Niche differentiation corresponds to different species which may be superior or inferior at specific functions in such a way that cumulatively they fare equally well and consequently co-exist.  In our model this simply corresponds to variable strategies and payoffs among players.  A competitive network model corresponding to niche differentiation which allows the competitive ability of individuals to be variable within each species therefore fits nicely into our theory.  

The number of co-existing species in a model based on niche differentiation without individual variation, such as \cite{al}, increases when competition is restricted to proximate neighbors.  In \cite{mi} local competition or temporal-niche differentiation is suggested as the underlying mechanism which supports the large biodiversity of planktonic micro-organisms.  That work is based on a lattice Lotka-Volterra predator-prey model which again sets a fixed competitive ability uniformly for all organisms of a single species.  The experimental data of  \cite{kerr} with bacteria of three fixed types showed that local competition which is periodically mixed does not result in coexistence (``mixed plate" experiment of \cite{kerr}).  Due to the dynamic and mixing environment inhabited by plankton, competition cannot remain local over long periods of time.  This laboratory experiment with three bacteria strains comprised of identical clones failed to show co-existence when bacteria were periodically mixed, however the numerical simulations of \cite{kerr} showed co-existence by incorporating individual variability by ``sequentially picking random focal points and probabilistically changing their states.''  Whereas localized competition may enhance species persistence, it is not a consistent mechanism to ensure survival if the species cannot express individual variation.  In contrast, individual variability was uniformly observed to promote coexistence in our spatially explicit RPS simulation.   This supports the notion that individual variability is perhaps the key mechanism supporting planktonic biodiversity.  

\subsection{Gause's Principle} 
Our prediction may appear to contradict Gause's competitive exclusion principle \cite{gause}.  However, Gause observed that competitive exclusion does not necessarily occur if ecological factors are \em variable. \em  We propose that the primary mechanism driving the large variability of ecological factors amongst plankton is individual variability.  When species possess variability across individuals, corresponding to the hypothesis in the theorem that each species has at least two pure strategies, then competition between individuals is unpredictable.  The success of the species is determined by the cumulative success across its individuals.  If a species is cumulatively inferior, our model predicts its extinction consistent with the competitive exclusion principle.  

If however species are on average cumulatively equally matched, then both the mathematical theory and our simulations predict neutral co-existence.  More importantly, there are \em infinitely many different \em probability distributions across individuals which have identical cumulative competitive abilities over time.  Each of these \em different \em probability distributions across individuals summarizes the cumulative characteristics of each \em different \em species.  The entirety of these diverse behavior distributions correspond to an unlimited biodiversity capable of neutrally co-existing.  
 
Our model does not imply that species will co-exist regardless of their environment.  Indeed, the payoff functions depend on the specific environmental conditions; we chose a simple payoff function for the purpose of simulation, but this is not required for our theorem.  If resources are scarce, then there may not be any SBDs, because the resources cannot support the entire population across all species.  However, even in this case, the mathematical theorem still applies in the sense that, if there is at least one strategy across all species with equal payoffs, then there are (generically) infinitely many such strategies.  This means that there are many different ways for all species to fare equally well, whatever the circumstances may be.  
 
Assuming equal cumulative competitive ability across all species with diverse behavior distributions may be inappropriate over short durations. We therefore tested the persistence of two species with unequal competitive abilities by generating SBDs for an inferior competitor with mean competitive ability 0.25 (Fig.~ \ref{fig2}). This inferior competitor was matched against a species with a bimodal SBD (Fig.~ \ref{fig2}) including weak (0-0.1) and strong (0.9-1.0) competitors. Thus, an inferior competitor could be competing against an individual with either poor or excellent competitive ability, and the outcome of competitions between these two species was not predictable \`a priori. Similar to the general case shown in Fig.~ \ref{fig3}, the inferior competitor persisted consistently when the populations size was $\gtrsim$ 3000 individuals. Even at very small population sizes of 10 individuals, the inferior competitor could persist for several generations. Thus, strategic behavior distributions help maintain species diversity including that of inferior species whereas competitive exclusion would predict rapid extinction.  

\subsection{Individual Variability}  
There is significant evidence for individual variability among planktonic micro-organisms.  For example, their spatial abundance has been modelled with biased random motion \cite{grun}.   Biased random motion implies an extremely high amount of variability in the motility patterns of different individuals within a single species.   If the environment is chaotically mixing, which is often the case for oceanic plankton, it is not possible to predict the optimal motility pattern and therefore a species expressing a variety of motility patterns will be more successful cumulatively over time than a species with a uniformly identical motility pattern.  Plasticity in a variety of traits, not only motility, and phenotypic diversity amongst planktonic micro-organisms has been demonstrated in \cite{hu01}, \cite{cell}, and \cite{Schaumetal2013}.  However, rather than expressing completely random motility patterns, \cite{kl} showed that the vertical motility patterns of phytoplankton are strategic to effectively optimize the acquisition of light and nutrients.  Phytoplankton may even flee from predators \cite{hmd} but this behavior is not observed to be uniform across all individuals. While motility patterns may appear to be somewhat random and variable when assessed at the individual level, they appear to be strategic on average over time when assessed cumulatively for a species.  Our model incorporates both individual variability and strategy.  A species need only be strategic as a whole, corresponding to one of the infinitely many SBDs, which allows unlimited variability across individuals.  However, an SBD assessed cumulatively over all individuals, while not necessarily an equilibrium strategy, is sufficient to ensure survival.  

A natural question is, why doesn't evolution drive species toward a Nash equilibrium strategy?   The crux of the paradox of the plankton seems to be that competition models which may be suitable for larger organisms do not allow sufficient flexibility to be consistent with the much greater biodiversity observed in plankton (and perhaps other) micro-organisms \cite{bald03}, \cite{bald08}, \cite{cuvel}, \cite{worsen}, \cite{massana}.  This is what lead us to seek a model which would allow a much larger set of strategies which are good enough to ensure survival and which allow a large variety of species to co-exist.  Viewing players as entire species allows for several \em different \em probability distributions across individuals each of which characterize a \em different \em species.  Based on feedback alone, there is no difference between all the different strategies in each level set of the payoff function.  

 \begin{figure}[ht] 
\begin{center}
\includegraphics[width=0.5\textwidth]{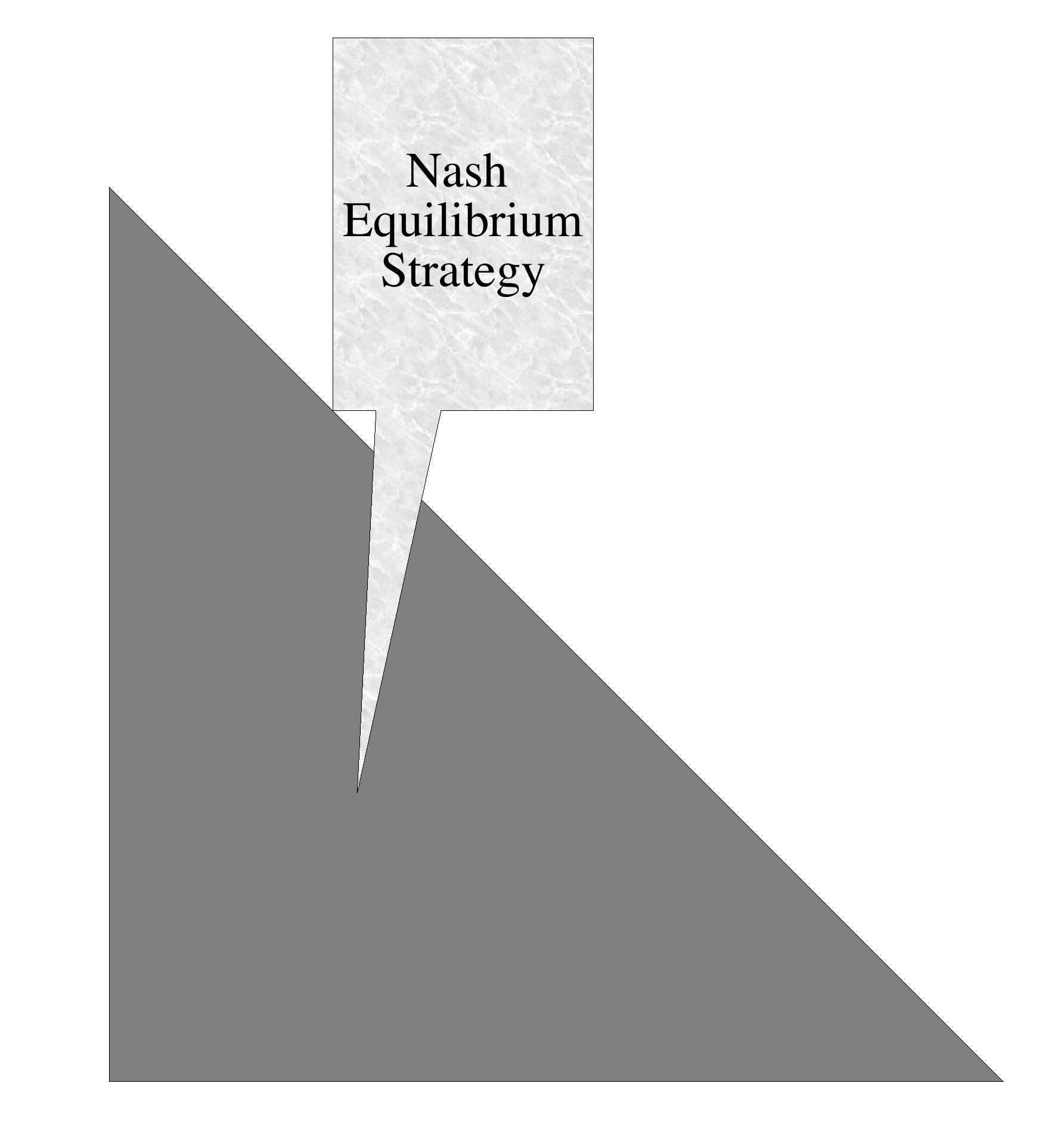}
\caption{For the RPS game, the set of SBDs can be represented by the set of all points inside an isosceles right triangle, whereas the Nash equilibrium strategy is a single point inside this triangle.} 
\label{rps} 
\end{center} 
\end{figure} 

The RPS game satisfies the hypotheses of our theorem because each player has at least three pure strategies.  Moreover, variations of that game such that it is no longer zero-sum also satisfy the hypotheses of our theorem.  For the classical RPS game, the unique Nash equilibrium strategy is the mixed strategy to draw Rock, Paper, and Scissors each with probability 1/3.  The corresponding payoff to players is 0.  The set of SBDs is the set of all mixed strategies such that the payoff to each player is zero, and in this case has dimension 2 and can be represented by the set of points in an isosceles right triangle.  One can imagine that the strategies of plankton species can vary -  including \em off \em of the set of SBDs - in such a way that their cumulative strategies roam around within this triangle or at least do so in a time-averaged sense.  This region, which contains infinitely many different SBDs, corresponds to the large biodiversity which can co-exist; see Fig. \ref{rps}.  The large set of SBDs in our mathematical model, not only for this example, but for any of the infinitely many games which satisfy the hypotheses of our theorem, allows for much greater biodiversity than a model which drives species toward a Nash equilibrium strategy.  

\subsection{Concluding Remarks} 
The novelty of our suggestion is to recognize the importance of individual-scale, cell-cell interactions that incorporate individual-level heterogeneity in behaviors and physiological characteristics, rather than assessing species, groups or aggregates based on average characteristics. Thus, the predictability of the outcome of competitions, which is near certain when assessed on the mean, becomes unpredictable when assessed on the individual level. This unpredictability results in the observed survival of individuals and persistence of diverse species. The mechanisms proposed here rely on species that are (1) highly diverse in one or many of their physiological, behavioral, or morphological functions, and (2) reproduce rapidly, asexually.  Competition is first assessed at an individual, organismal level and then cumulatively assessed over all individuals to determine the resulting outcome for the species as a whole. 

Our model for the population dynamics likely only applies to \em microbial organisms \em with large population sizes, but we have not explored the issue systematically.   For clonal organisms, each individual only represents a small fraction of the total population, and thus variable, and probable map-adaptive behaviors pose no risk to the overall gene pool. For sexually reproducing organisms, each individual is unique (although there may be great similarities among groups). Thus, we interpret our model to be a poor fit for species where each organism represents a unique set of evolved characteristics that would be lost to the population if the individual were to perish. This risk is not inherent in clonal populations.

Our model possesses the strength that it is insensitive to specific formulations; payoffs for strategies can be constant or varying among species or over time, and the distribution of functions within a species can be constant or varying over time. The key factor is that in microbial systems each individual represents only a small fraction of the total population, where survival of each individual is associated only with a small risk. Consequently, some individuals can represent extreme strategies that may not be \`a priori suitable or advantageous but may be successful infrequently. 

\section{Methods}
\begin{figure}
        \centering
        \begin{subfigure}[b]{0.3\textwidth}
                \includegraphics[width=\textwidth]{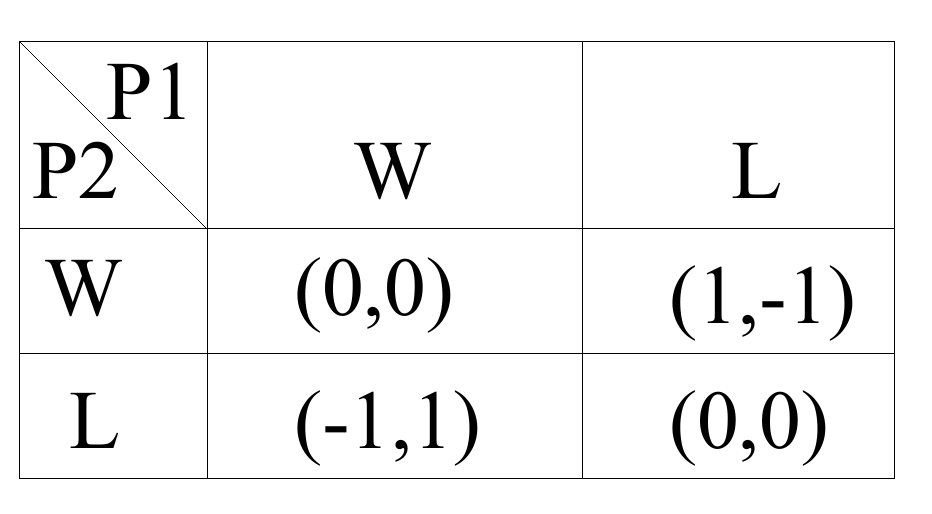} 
                \caption{}
                \label{avsb}
        \end{subfigure}%

        \begin{subfigure}[b]{0.4\textwidth}
                \includegraphics[width=\textwidth]{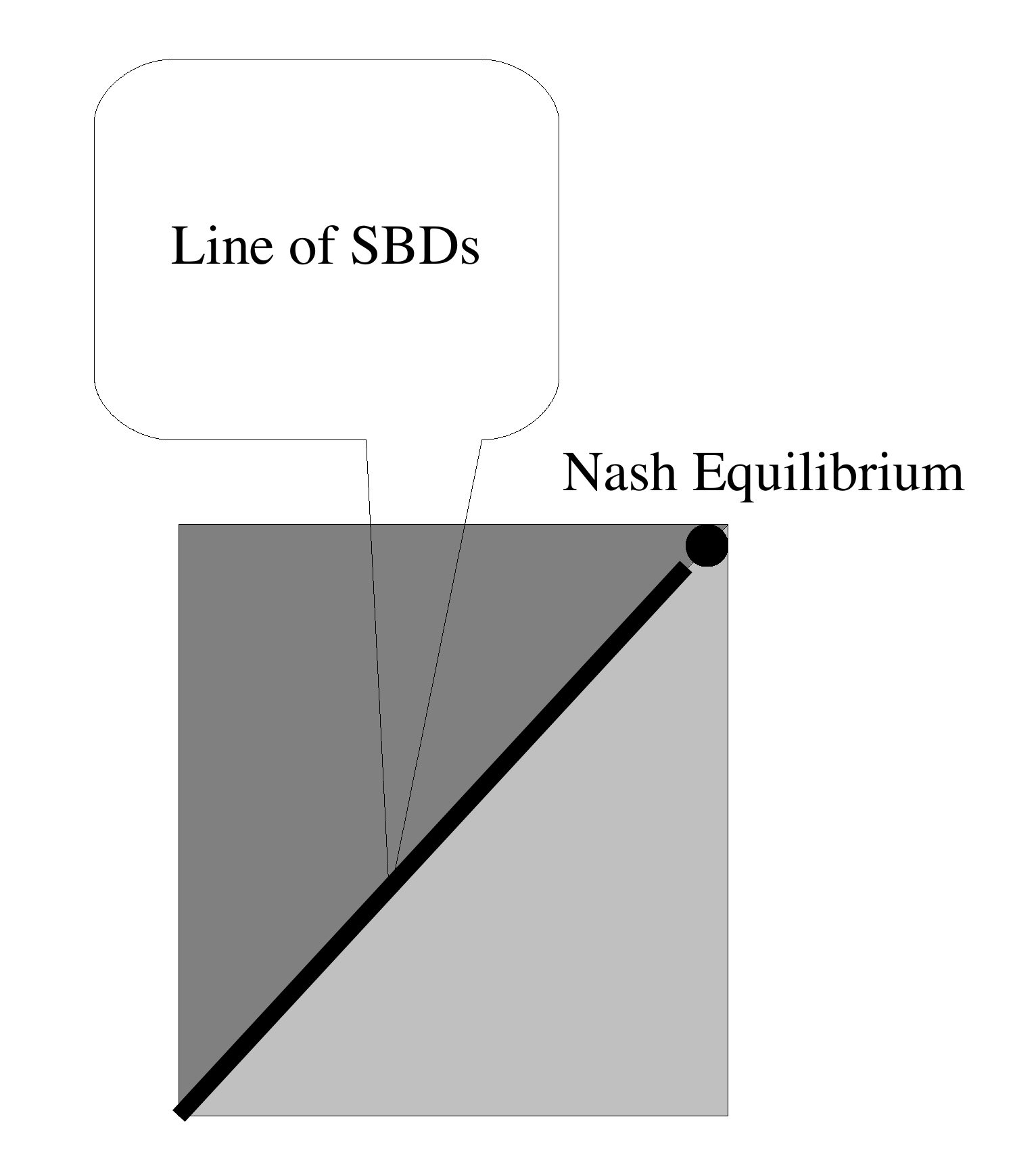}
                \caption{}
                \label{sbd-wl}
        \end{subfigure}
        \begin{subfigure}[b]{0.32\textwidth}
                \includegraphics[width=\textwidth]{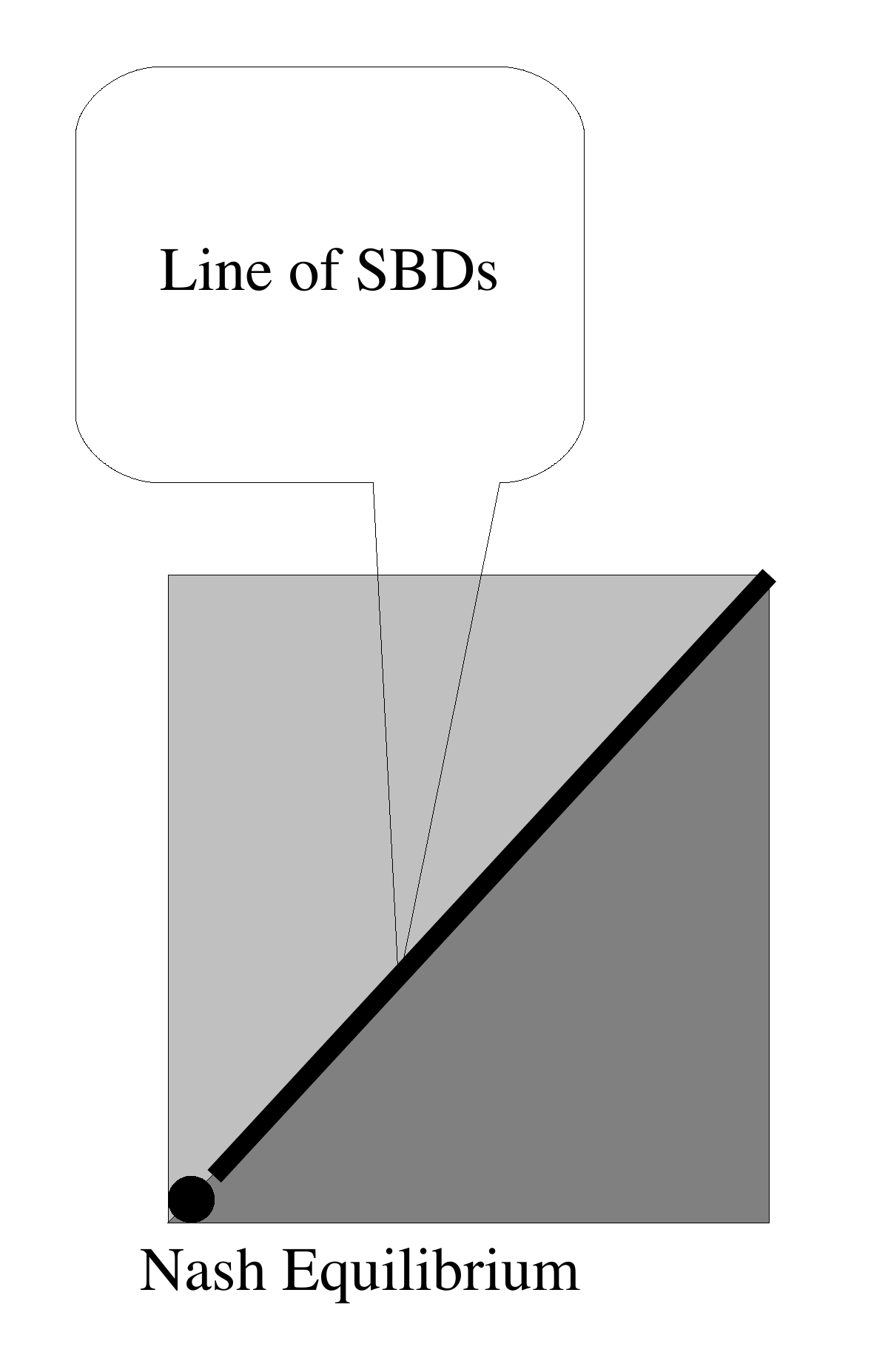}
                \caption{}
                \label{fig:mouse}
        \end{subfigure}
        \caption{(A) shows the payoff matrix for the symmetric, two-player, zero-sum game used to simulate competition.  (B) shows the line of SBDs and the unique Nash equilibrium strategy.  Player 1 has a higher payoff in the darker region, and player 2 has higher payoff in the lighter region; both players have identical payoffs along the line of SBDs and at the Nash equilibrium strategy.  (C) shows what happens if the dominant strategy changes.   The lighter and darker regions exchange places, and the Nash equilibrium point is at the other end of the line of SBDs.  The SBDs however, remain unchanged. }\label{fig1}
\end{figure}

\begin{figure}
        \centering
        \begin{subfigure}[b]{0.4\textwidth}
                \includegraphics[width=\textwidth]{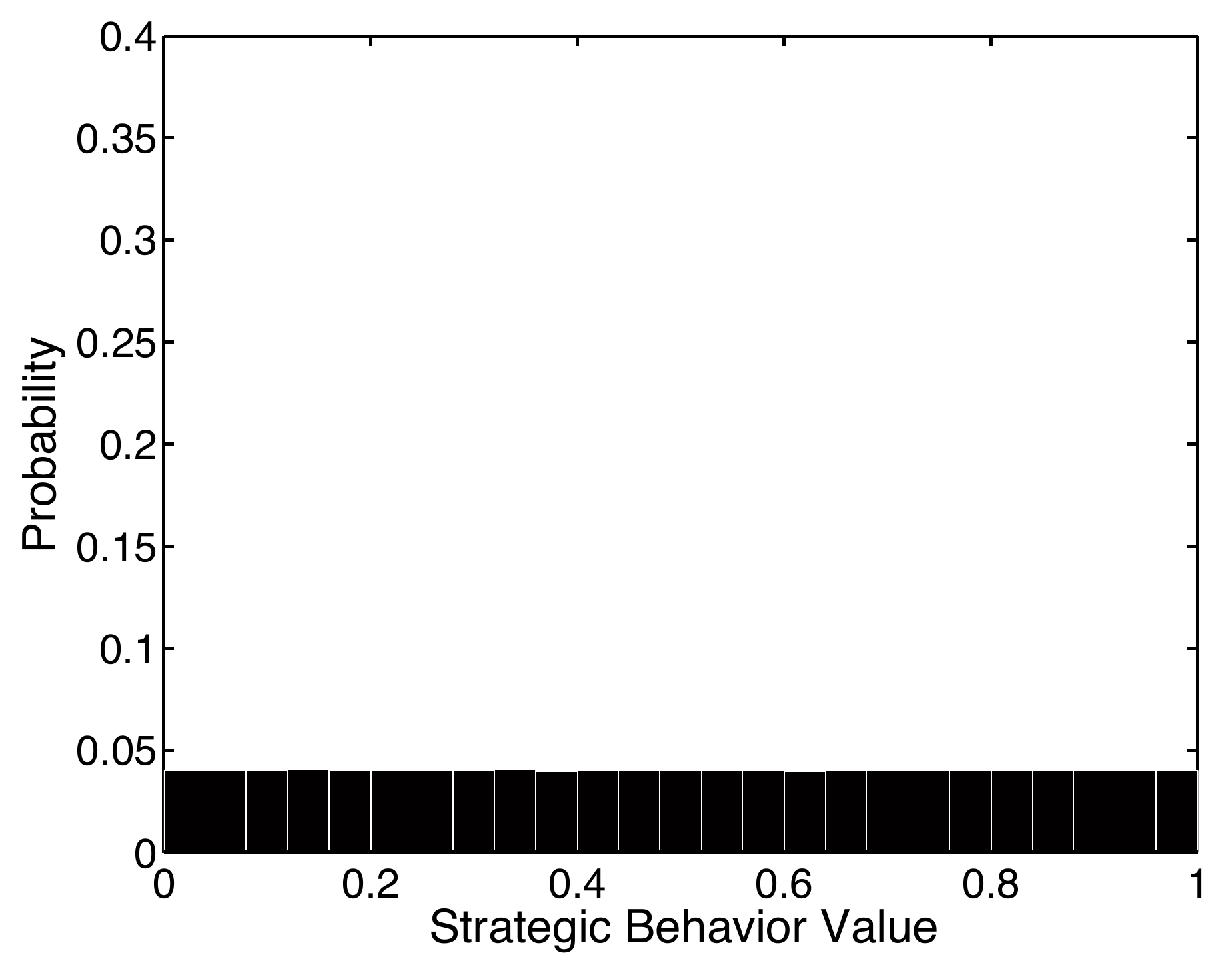}      
                \caption{}
                \label{whole}
        \end{subfigure}%
        \begin{subfigure}[b]{0.4\textwidth}
                \includegraphics[width=\textwidth]{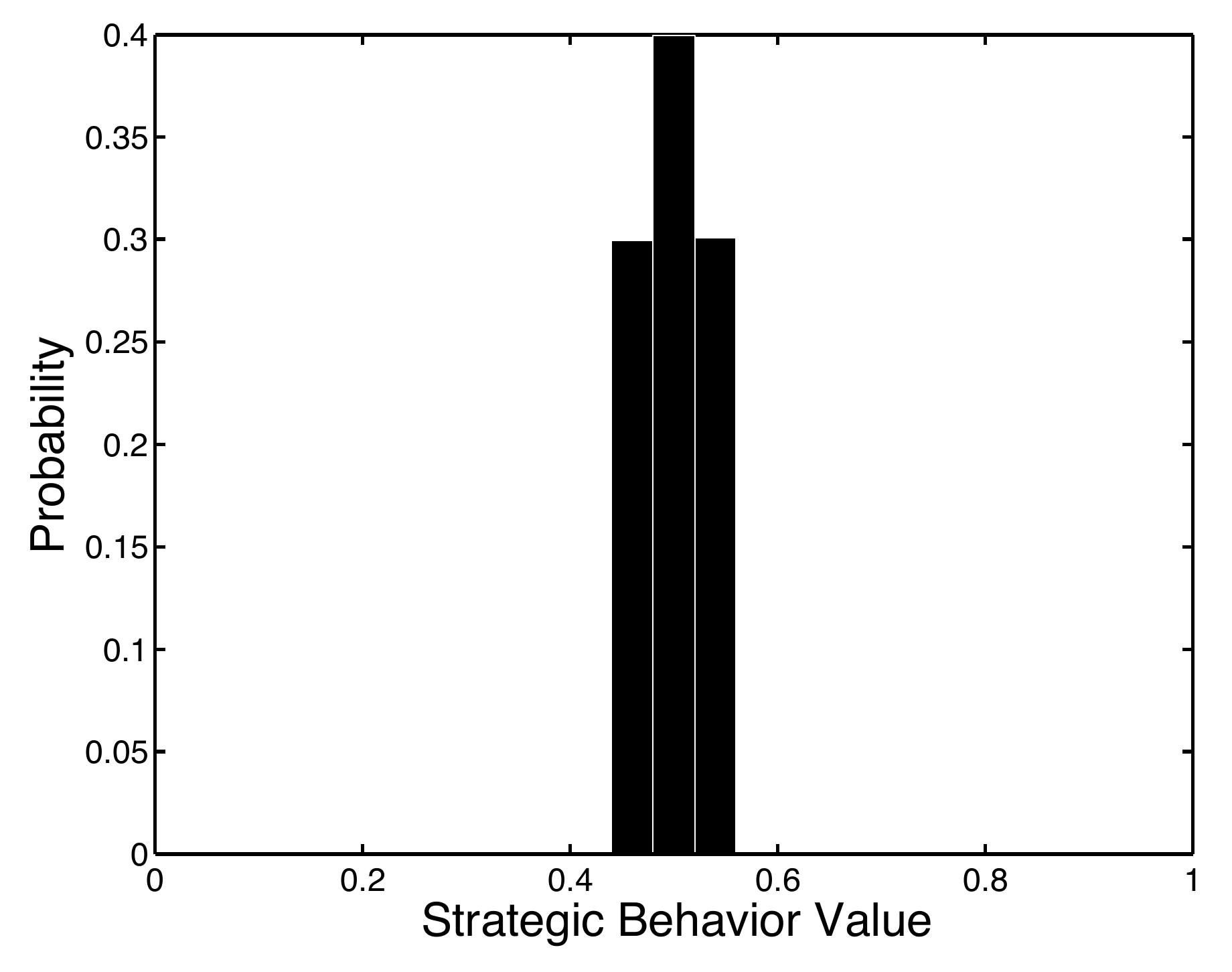}
                \caption{}
                \label{center}
        \end{subfigure}
        
        \begin{subfigure}[b]{0.4\textwidth}
                \includegraphics[width=\textwidth]{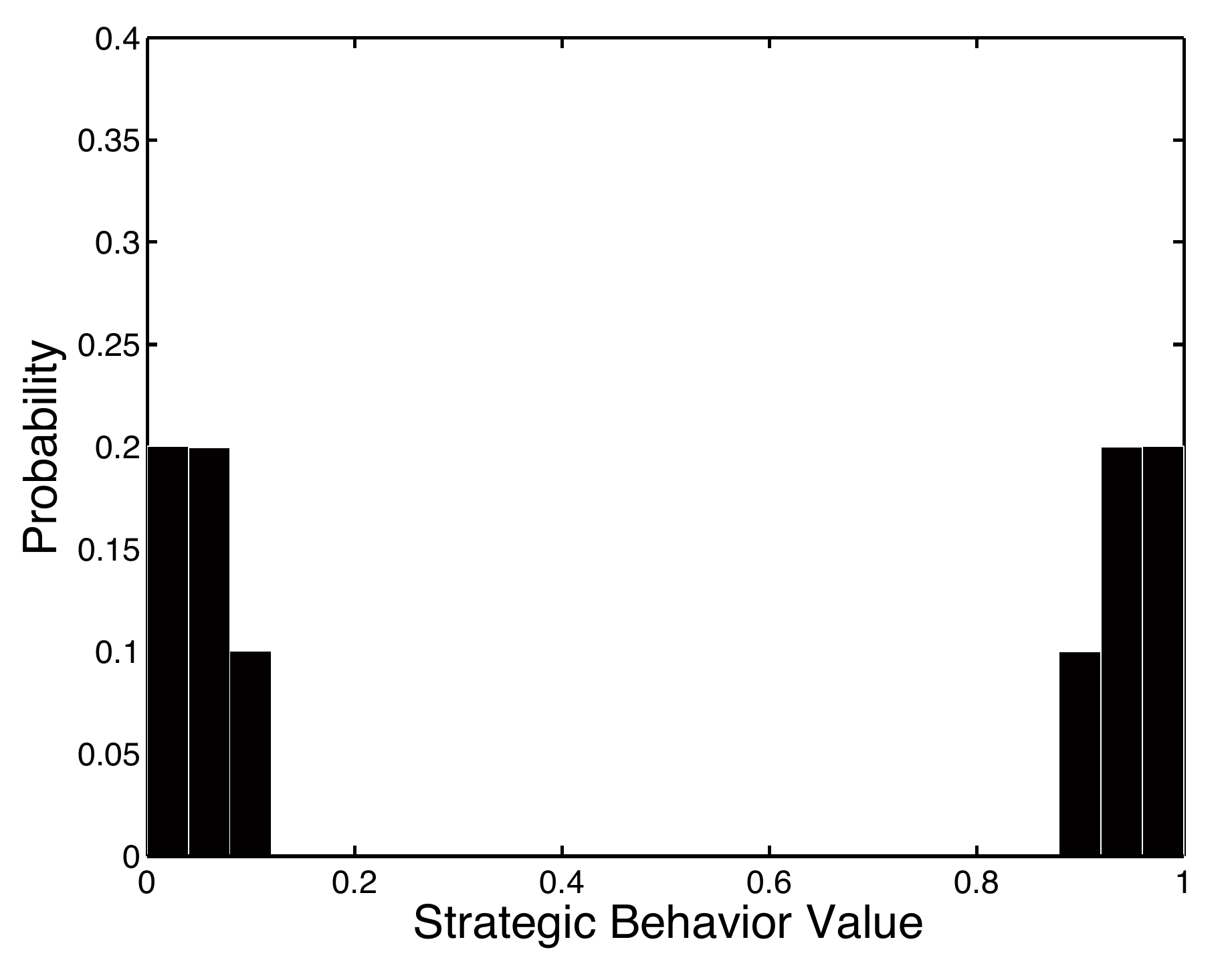}
                \caption{}
                \label{edges}
        \end{subfigure}
                \begin{subfigure}[b]{0.4\textwidth}
                \includegraphics[width=\textwidth]{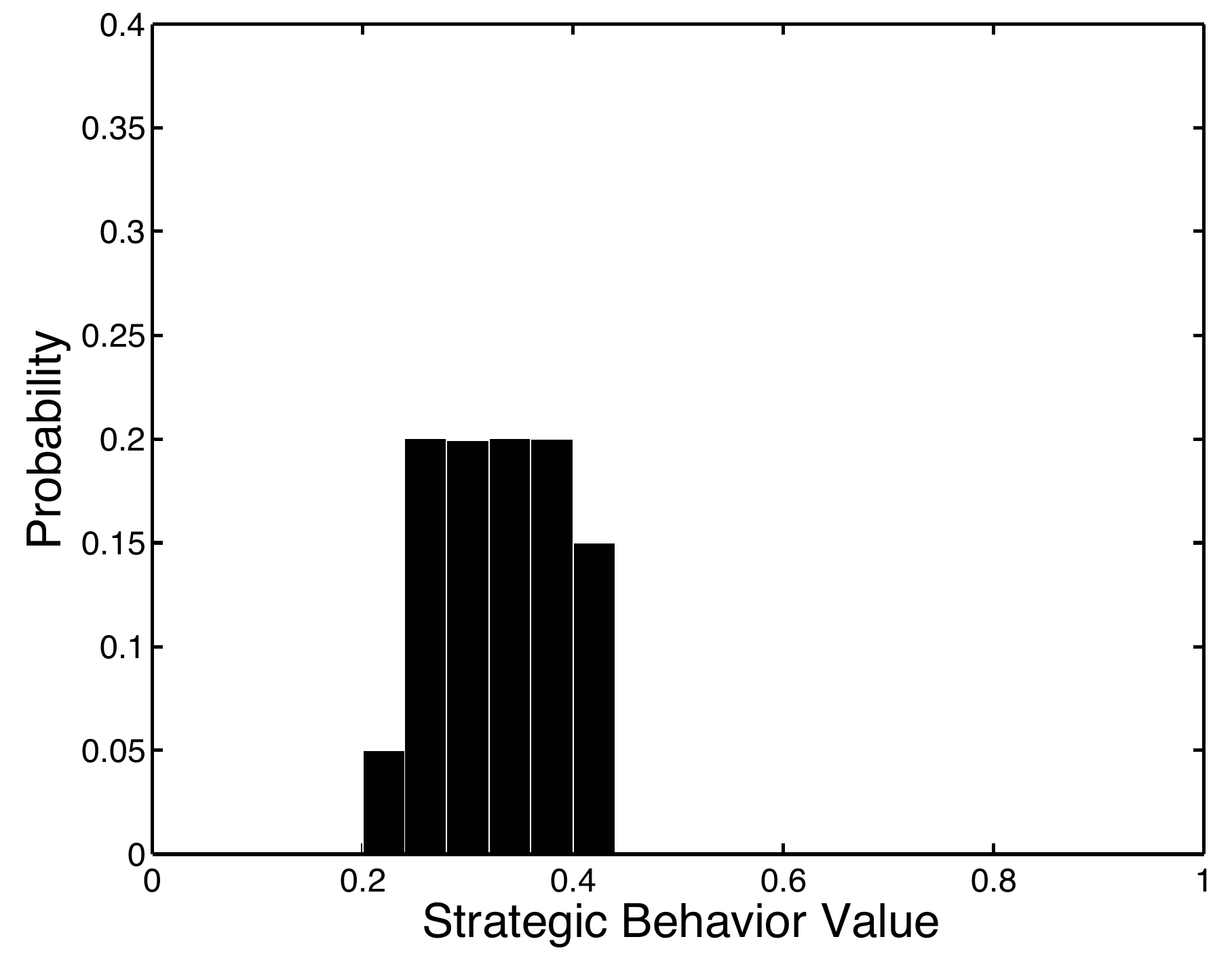} 
                \caption{}
                \label{inferior}
        \end{subfigure}
    \caption{(A), (B), and (C) show the non-normal SBDs all with mean 0.5 from which individuals' competitive abilities were randomly drawn to implement the competition simulations.  (D) shows the SBD for an inferior species with mean competitive ability 0.25.}
\label{fig2} 
        \end{figure}

We arrived at a key discovery based on the following simple example.  Consider a symmetric, two-player, zero-sum game with two pure strategies, W and L, such that W (win) dominates L (lose) (see Fig.~ \ref{fig1}). If the probability that players 1 and 2 execute strategy W are respectively $s_1$ and $s_2$, then the unique equilibrium strategy is $s_1=s_2 =1$, and the payoff to both players is 0.   \em There are infinitely many strategies such that the payoff to each player is identical to their payoff according to the equilibrium strategy.  \em  The payoff functions are $\wp_1 (s_1, s_2) = s_1 - s_2,$ and $\wp_2 (s_1, s_2) = s_2 - s_1$.  The set of SBDs is the line segment $\{0 \leq s_1 = s_2 \leq 1\}$  If the game changes so that strategy L dominates strategy W, the Nash equilibrium becomes $(s_1=s_2=0)$, but the set of SBDs remains unchanged (see Fig.~ \ref{fig1}).  Although an SBD is not optimal in the sense of a Nash equilibrium, in this example it gives \em identical feedback \em to the Nash equilibrium.  This example illustrates our mathematical theorem.  In this case the total payoff function is a linear function, and since the game is zero-sum, $\wp_2 = - \wp_1$.  Therefore, the total payoff function can be canonically identified with a linear map from $\R^2 \to \R$, and the level sets of this function have dimension one (the level sets are line segments).  This shows that for each value of the payoff functions (i.e. feedback to each competing species), there is a whole \em line \em of strategies with those same payoff to each respective player.  

\subsection{Proof of the Mathematical Theorem} 
The main ingredient in the proof of our theorem is the co-area formula in geometric measure theory due to Federer \cite{fed}.  One of the implications of this theorem is that a Lipschitz continuous function from a higher dimensional Euclidean space, like $\R^m$ into a lower dimensional Euclidean space, like $\R^l,$ such that $l < m$ generically has large level sets.  More precisely, the rank of the differential of such a function $f$ is with respect to $m-l$ dimensional Hausdorff measure almost-always equal to $l$.  Consequently, the level set $f^{-1} (f(x))$ contains a subset which is a topological submanifold of $\R^m$ of dimension $m-l$.  In the more simple case of an affine linear function as in Figure \ref{fig1},  the dimension of the level sets is given by the rank-nullity theorem from linear algebra.  For an affine linear function $f$ from $\R^m \to \R^l$, there is $y \in \R^l$ and an $l \times m$ matrix $M$ such that for all $x \in \R^m$,  $f(x) = Mx + y$.  Consequently the level sets are all affine linear subsets of $\R^m$ with dimension equal to the dimension of the kernel of $M$.  Since $M$ is $l \times m$ and $m>l$, by the rank-nullity theorem the dimension of this kernel is at least $m-l$.  

Under the hypotheses of the theorem, the total payoff function $\wp = (\wp_1, \wp_2, \ldots, \wp_n)$ is canonically identified with a map from a convex $N-n$ dimensional subset of $\R^N$ into $\R^n$.  This is because the sum of all components of a strategy must be equal to one, and hence the last component of each player's mixed strategy is determined by the previous components.  Given that there are $n$ players total, this means that the payoffs are determined by elements of $\R^{N-n}$.  Since all players have at least two pure strategies, $N\geq 2n$ and so $N-n \geq n$.  If the game is zero-sum, then 
$$\wp_n = - \sum_{k=1} ^{n-1} \wp_k,$$
and hence the total payoff function is canonically identified with a map from $\R^{N-n}$ into $\R^{n-1}$, and in this case $N-n \geq n > n-1$, and we define $k:=N-2n +1$.  If only (2) holds, then $N > 2n$, and thus $N-n > n$, and the total payoff is canonically identified with a map from a convex $N-n$ dimensional subset of $\R^N$ into $\R^n$.  In this case $k:=N-2n$.  Consequently, in all cases the total payoff function is canonically identified with a map from a larger dimensional Euclidean space $\R^m$ for $m=N-n$ into strictly smaller dimensional Euclidean space $R^l$ for $l= m-k$.  By the co-area formula the level sets of the total payoff function with respect to $k$-dimensional Hausdorff measure almost always have positive $k$-dimensional Hausdorff measure.  
\qed

\subsection{Statistical Mean of Tournament Simulations} In each simulation, a species was represented by at least one and up to 10000 individuals.  Each tournament included several randomization parameters, including the random assignment of behavior values to individuals as well as randomizing the selection of competitors. To explore the robustness of these results, we varied the number of individuals per species in each type of simulation over the entire range from 1 to 10000. To eliminate the effect of this stochastic variation, we repeated each simulation of a particular parameter combination at least 30 times and computed the mean over all runs.  All simulations were run in Matlab R7.10.

\subsection{Spatially Explicit Model} We included spatial information in the tournament by formulating a lattice structure in which each individual was initially assigned a species, strategic behavior, and location at random. In this spatially explicit model we additionally examined the role of local versus global competition, which has been found to be an important structuring factor \cite{kerr}. The boundary conditions were periodic for the spatially explicit model. The time step of evaluation (t) was repeated up to 10000 times. The population abundance of each species and if spatially explicit, location, of each individual were determined at each time step.  All simulations were run in Matlab R7.10.

\section*{Acknowledgments}
We thank Henry Segerman and Mick Follows for critical reading of this manuscript and the anonymous reviewers whose comments greatly improved this paper.  We gratefully acknowledge the support of the National Science Foundation (Biological-Oceanography Award 0826205 to S. Menden-Deuer), the Max Planck Institut f\"ur Mathematik, the Universit\"at G\"ottingen, the Leibniz Universit\"at Hannover, and the Australian National University.

\end{document}